\documentclass[conference]{IEEEtran}
\usepackage{amsmath,amssymb,amsfonts,amsthm,mathrsfs,bm}
\usepackage{graphicx}
\usepackage{balance}
\usepackage{cite}
\usepackage{graphics}
\usepackage{graphicx}
\usepackage{subfig}
\usepackage{epsfig}
\usepackage{caption}
\usepackage{algorithm}               
\usepackage{algorithmic}             
\usepackage{multirow}                
\usepackage{amsmath}
\usepackage{xcolor}

\IEEEoverridecommandlockouts
\newtheorem{theorem}{Theorem}
\newtheorem{lemma}{Lemma}
\theoremstyle{definition}

\theoremstyle{remark}

\theoremstyle{corollary}

\theoremstyle{definition}
\theoremstyle{Definition}
\allowdisplaybreaks

\begin{document}
\title{Delay Optimal Scheduling of Arbitrarily Bursty Traffic over Multi-State Time-Varying Channels}
\author{
\IEEEauthorblockN{Meng Wang, \IEEEmembership{Student Member, IEEE}, Juan Liu, Wei Chen, \IEEEmembership{Senior Member, IEEE}\\}
\IEEEauthorblockA{
Tsinghua National Laboratory for Information Science and Technology (TNList)\\
Department of Electronic Engineering, Tsinghua University, Beijing, 100084, CHINA\\
Email: m-wang14@mails.tsinghua.edu.cn, eeliujuan@gmail.com, wchen@tsinghua.edu.cn}
}
\maketitle
\begin{abstract}
In this paper, we study joint queue-aware and channel-aware scheduling of arbitrarily bursty traffic over multi-state time-varying channels, where the bursty packet arrival in the network layer, the backlogged queue in the data link layer, and the power adaptive transmission with fixed modulation in the physical layer are jointly considered from a cross-layer perspective. To achieve minimum queueing delay given a power constraint, a probabilistic cross-layer scheduling policy is proposed, and characterized by a Markov chain model. To describe the delay-power tradeoff, we formulate a non-linear optimization problem, which however is very challenging to solve. To handle with this issue, we convert the optimization problem into an equivalent Linear Programming (LP) problem, which allows us to obtain the optimal threshold-based scheduling policy with an optimal threshold imposed on the queue length in accordance with each channel state.
\end{abstract}

\section{Introduction}

Qualities of Service (QoS) such as low latency is expected in next generation wireless communication system to provide real time multimedia services, such as H.265 video streaming\cite{6824752}. It is a key metric to measure the quality of serving delay-sensitive or time-critical applications. In the meantime, high energy efficiency is urgently required especially for mobile terminals which are usually powered by rechargeable batteries of finite capacities in wireless systems. However, there is a fundamental tradeoff between the average queueing delay and the average power/energy consumption. Intuitively, to reduce the latency, the transmitter would conduct transmission more frequently, which however would increase the probability of suffering bad channel states. Thus, much more power will be consumed. 

In general, it is very challenging to derive the delay-power tradeoff in wireless communication systems, considering the randomness of data packet arrivals, and the time-varying characteristics of wireless channels. In the last decades, the cross-layer design framework was proposed to deal with the uncertainties occurring at different layers. It also provided an effective way to measure the average queuing delay in the data link layer and the average power consumption in the physical layer, and then reveal the optimal delay-power tradeoff in various system settings. To our best knowledge, the idea of jointly combining the network-layer data arrival and the physical-layer data transmission was firstly presented in \cite{collins1999transmission}, where the average transmission power is minimized under the constraints of peak power and average delay. In \cite{995554}, under the framework of cross-layer, Berry and Gallager aimed to regulate the average transmission power and average buffer delay by adapting users' transmission power and rate. Then, several basic cross-layer resource allocation problems, such as power allocation and rate adaption, were studied for wireless fading channels in \cite{berry2004cross}. Ata focused on the power minimization problem subject to the packet drop rate in \cite{ata2005dynamic} under the assumption of fixed channel state, Poisson packet arrival and exponentially distributed packet size. In the past years, a lot of works on the delay-power tradeoff analysis have been done by applying the cross-layer design approach.

In existing works, different optimization methods have been applied to achieve the optimal delay-power tradeoff. Network calculus was used to model energy-efficient transmission with deadline constraint in \cite{uysal2002energy}, \cite{zafer2009calculus}. The authors proposed both offline and online algorithms named Lazy scheduling algorithms to minimize the transmission power given deadline constraint in \cite{uysal2002energy}. Based on cumulative curves methodology, the optimal transmission policy for minimizing the transmission power under the QoS constraints was obtained in \cite{zafer2009calculus}. Dynamic Programming (DP) algorithm was used to derive the optimal scheduling policy in \cite{collins1999transmission}, \cite{995554}, \cite{4533700}. 
Besides, constrained Markov decision processes (CMDP) \cite{altman1999constrained}, \cite{hernandez2012discrete} was adopted to formulated the delay and power tradeoff. In \cite{4567575}, with CMDP, the authors obtained the optimal policy to achieve the power and delay tradeoff for the considered single-user system and multi-user system.

More recently, we focus on studying the delay-power tradeoff in wireless transmissions from the perspective of cross-layer probabilistic scheduling.  
A probabilistic scheduling policy was proposed to achieve the minimum queueing delay under the constraint of transmission power in our previous work \cite{chen2007optimal}, where Bernoulli distributed packet arrivals and a two-state fading channel model were considered. To capture the influence of bursty network traffic, we studied the delay-power tradeoff in wireless systems with arbitrarily random packet arrival patterns in \cite{7417380}. In these works, we proved that the optimal delay-power tradeoff can be achieved by applying the optimal scheduling polices which determine packet transmissions based on the optimal thresholds imposed on the queue length. In this paper, we study the delay-power tradeoff in wireless packet transmissions of bursty traffic over multi-state wireless channels. In contrast to our previous works\cite{chen2007optimal}\cite{7417380}, we propose a probabilistic cross-layer scheduling policy which schedules packet transmissions based not only on the queue length but also on the channel state. Using  Markov reward process theory, we derive the expressions of the average delay and the average power, and formulate an optimization problem to describe the delay-power tradeoff. We show that the optimal scheduling policy still has the threshold structure. The major difference between this work and existing works lies in that the optimal threshold relies both on the number of backlogged data packets and on the channel state. Before detailed discussion, we introduce two symbols as $a \wedge b=max\{a,b\}$ and $a\vee b=min\{a,b\}$. Throughout the paper, the proofs of lemmas and theorems are omitted due to limited space.


\section{System Model} \label{sec2}

\begin{figure}[t]
\centering
\captionsetup[subfloat]{labelformat=simple,captionskip=6bp,nearskip=6bp,farskip=0bp,topadjust=0bp}
\renewcommand{\thesubfigure}{(\alph{subfigure})}
\subfloat[Wireless Link]{
\includegraphics[width=0.8\columnwidth]{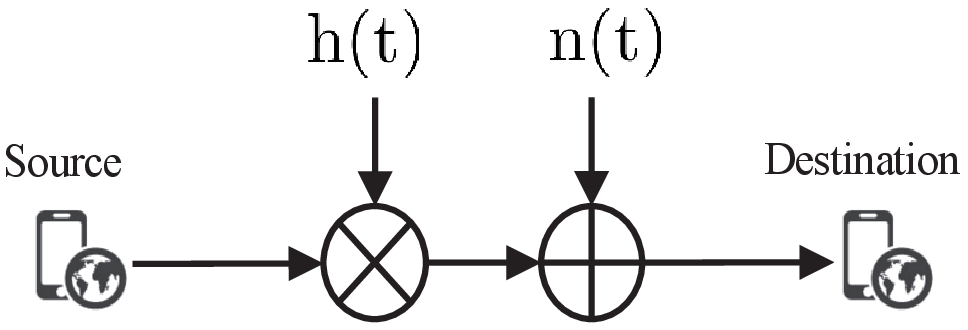} \label{model_1}}\\
\subfloat[Cross Layer System Model]{
\includegraphics[width=1\columnwidth]{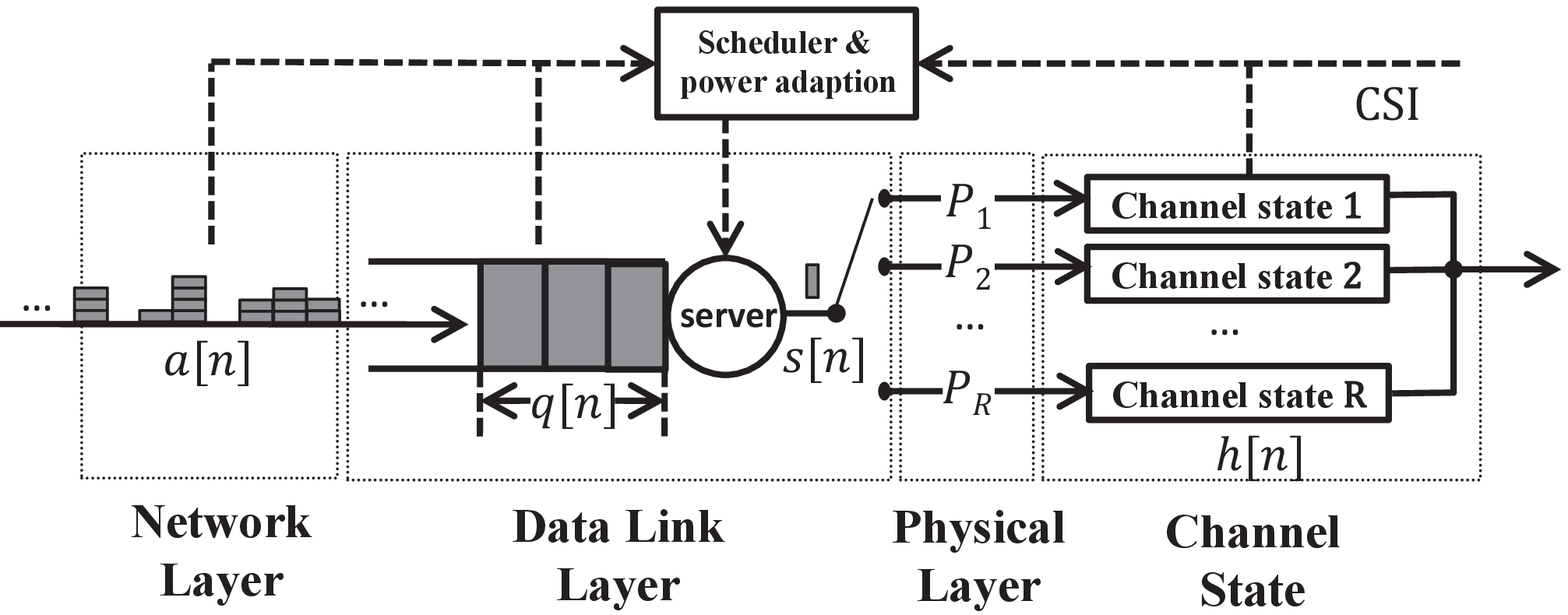} \label{model_2} }
\caption{System Model}
\label{model}
\vspace{-0.0cm}
\end{figure}

As shown in \figurename\ref{model_1}, we consider a wireless communication system where the source node transmits to the destination over a time-varying  wireless link. We will introduce the system model in a cross-layer design framework, i.e., packet arrivals of bursty traffic in the network layer, queueing behaviour in the data link layer, power control and wireless data transmission in the physical layer, as shown in \figurename\ref{model_2}.

Considering bursty traffic, data packets generated by higher-layer applications arrive at the network layer randomly. Let $a[n]$ denote the number of packets randomly arriving in the $n$th slot. To capture the burstiness and variability of real-time applications, we assume an arbitrarily packet arrival pattern, i.e., the number of newly arriving packets could follow any distribution. $a[n]$ is assumed to follow an independent and identically distributed ($i.i.d.$) process. Thus, the mass probability function of $a[n]$ can be characterized by
\begin{align}\label{eq1}
\text{Pr}\{a[n]=m\}=\theta_m,  \  \  \  m = 0,1,2,\cdots.
\end{align}
where $\theta_m \in [0,1]$. Corresponding to traffic shaping and admission control adopted in the system, the number of packets newly arriving in each time slot must be upper-bounded by a large integer $M$, i.e., $\exists \  M \geqslant 0, \forall \ m > M,\  \theta_m = 0$. Since the distribution of $a[n]$ shall be properly normalized, we have $\sum_{m=0}^{M} \theta_m=1$. The average packet arrival rate $\bar{a}$ is obtained as
\begin{align}\label{eq2}
\bar{a}=\lim\limits_{N\rightarrow\infty}sup \ \frac{1}{N}\sum\limits_{n=0}^N a[n] =\sum_{m=0}^M m\cdot \theta_m.
\end{align}

At the source node, a buffer is employed to store the packets which can not be sent immediately. Without loss of generality, we assume that the buffer capacity $K$ is sufficiently large and buffer overflow could be negligible. The queue state, denoted by $q[n]$, is characterized by the number of packets in the buffer at the end of $n$th slot. It is updated as
\begin{equation}\label{eq3}
\begin{split}
q[n]&=max\big\{ min\{q[n\!-\!1]+a[n], K\}-s[n] , 0 \big\}\\
\end{split}
\end{equation}
where $s[n]$ denotes the number of packets delivered in the $n$th time slot.

We adopt a $W$-state block fading channel model, where $W$ is a positive integer. That is, the channel state stays invariant during each time slot and follows an $i.i.d.$ fading process across the time slots. In each slot, the channel stays in one of $W$ states according to the current channel condition. Let $d_1 = \infty > d_2 >\cdots > d_W > d_{W+1} = 0$ be the channel gain levels. If the channel gain in the $n$th time slot ranges in interval $[d_w,d_{w+1})$, we can say that the wireless channel is at $'Channel \  state \  w'$. 
One can see that, the channel quality becomes worse with the increase of the index $w$. Hence, $'Channel \  state \  1\!'$ and $'Channel \  state \  W'$ represent the best and the worst channel condition, respectively. 
 The mass probability function of the random variable $h[n]$ is described as
\begin{align}\label{eq4}
\text{Pr}\big\{h[n] = 'state\  w' \big \}=\eta_w,
\end{align}
where $\eta_w \in [0,1]$ and $ w\in \{1,2,\cdots,W\}$.

\begin{figure*}[t]
\centering
\includegraphics[width=2\columnwidth]{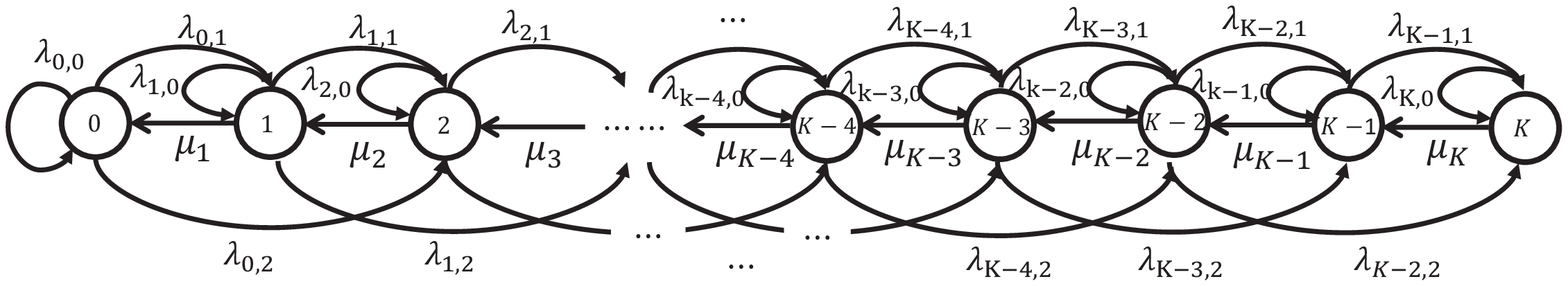}
\caption{The illustrative Markov chain model with $M=2$}
\label{markov}
\vspace{-0.2cm}
\end{figure*}

Suppose that there exits a feedback channel through which the Channel State Information (CSI) is sent back from the receiver to the transmitter. Intuitively, the transmission power shall be adapted to the channel state to meet the bit error rate (BER) requirement at the receiver side. Let $P_w$ ($1\leqslant w \leqslant W$) denote the power needed to transmit one packet successfully in the channel sate $w$. Since more power is required to combat wireless channel fading when the channel condition is worse, it is reasonable to assume $P_1<P_2<\cdots<P_w<\cdots<P_W$.

In our model, we consider fixed-rate transmission schemes which have been widely adopted in practice \cite{qiao2009impact}. Without loss of generality, we assume the transmission rate is one packet per slot. Hence, there is at most one data packet can be delivered in each slot, i.e., $s[n]\in\{0,1\}$.

In the cross-layer design framework shown in \figurename\ref{model_2}, the scheduler will schedule data transmissions based on the data arrival state $a[n]$, the queueing state $q[n-1]$, and the channel state $h[n]$, as will discussed in details in the next section.

\section{Probabilistic Scheduling and Markov chain Model}\label{sec3}

In this section, we will introduce a probabilistic scheduling policy and establish a discrete-time one-dimensional Markov chain to model the queueing system.

\subsection{Probabilistic Scheduling}\label{scheduling policy}

To improve power efficiency, the transmitter should exploit a better channel state to deliver the packets since it will spend much less power. Thus, the source is more willing to keep silent till the channel state gets better. However, this may induce undesirable large latency to wait for good channel states, which is intolerable for serving delay-sensitive or time-critical traffics. To overcome this, some backlogged packets should be transmitted immediately at the cost of higher power consumption, even when the channel state may not be so good. Hence, the proposed scheduler would be designed to achieve the balance between the average delay and the average power.

In this work, a probabilistic cross-layer scheduling policy is proposed.  
We assume that, at the beginning of the $n$th time slot, the channel state $h[n]='state\  w'$, the queueing state $q[n-1]=k$, and the data arrival state $a[n]=m$. The scheduler decides to transmit one packet with probability $f_{\{k,m,w\}}$ or keep silent with probability $1-f_{\{k,m,w\}}$. Assume that one of the packets newly arriving at this slot can be delivered immediately. In our system, it is not necessary to distinguish between the backlogged packets and the newly arriving packets. In this sense, the source will transmit one packet with probability $f_{k+m,w} \in [0,1]$ based on the total number of packets in the buffer after one packet arrival, $q[n]=k+m$ and the channel state, $h[n]='state\  w'$. Given $q[n-1]=k$, $a[n]=m$, $h[n]='state\  w'$, the scheduling policy can be described as
\begin{align}\label{eq5}
s[n]=\left\{
\begin{array}{ll}
1  \quad w.p. \quad f_{k+m,w},\\
0  \quad w.p. \quad 1-f_{k+m,w},
\end{array}
\right.
\end{align}
where $0 \leqslant k \leqslant K$, $0 \leqslant m \leqslant M$, $1\leqslant w \leqslant W$ and the abbreviation $'w.p.'$ is short for $'with\  probability'$. In Eq. \eqref{eq5}, when $a[n]=0$ and $q[n-1]=0$, there is no packet waiting to be transmitted while when $a[n]+q[n-1]>K$, packet loss will happen, thus $f_{0,w}$ ($1\leqslant w \leqslant W$) and $f_{k+m,w}$ ($k+m>K,\ 1\leqslant w \leqslant W$) are set as zero for notational consistence.

The proposed scheduling policy totally depends on a set of probabilities $\{f_{k+m,w} \ |\ 0 \leqslant k \leqslant K, 0 \leqslant m \leqslant M, 1\leqslant w \leqslant W\}$. The optimal delay-power tradeoff can be achieved by optimizing the probabilistic parameters ${f_{k+m,w}}$.

\subsection{Markov Chain Model} \label{steady-proc}

Based on the description of the scheduling policy in Sub-section \ref{scheduling policy}, a queueing system is modelled as a discrete-time Markov chain, where each state represents the buffer state $q[n]$. Let $\tau_{k,l}$ denote the one-step state transition probability from state $q[n-1]=k$ to state $q[n]=l$, i.e., 
\begin{equation}
\begin{array}{l}
Pr\{q[n]=l|q[n\!-\!1]=k,q[n\!-\!2]=k_1\cdot\cdot\cdot \} \\=Pr\{q[n]=l|q[n\!-\!1]=k\}\\=\tau_{k,l}\nonumber.
\end{array}
\end{equation}
Combining Eq. \eqref{eq3} and the ranges of variables $a[n]$ and $s[n]$, we know that, the transition probability $\tau_{k,l}$ satisfies the constraint that $\tau_{k,l}=0$ for $|k\!-\!l|>M$.

\begin{theorem}\label{theorem1}
The transition probabilities denoted by $\lambda_{k,m}=\tau_{k,k+m}$ and $\mu_k=\tau_{k,k-1}$ are obtained by
\vspace{-0.0cm}
\begin{align}
\lambda_{k,m}\!& = \theta_m\sum\limits_{w=1}^W \eta_w(1-f_{k\!+\!m,w}) + \theta_{m\!+\!1}\sum\limits_{w=1}^W \eta_w f_{k\!+\!m\!+\!1,w} ,\label{eq601}
\end{align}
\vspace{-0.0cm}
where $0 \leqslant k \leqslant K$ and $1 \leqslant m \leqslant M$,
\vspace{-0.1cm}
\begin{align}
\mu_k& = \theta_0\sum\limits_{w=1}^W \eta_w f_{k,w},\label{eq602}
\end{align}
\vspace{-0.0cm}
where $1 \leqslant k \leqslant K$. $\lambda_{k,0}$ is given by
\vspace{-0.0cm}
\begin{align}\label{eqk0}
\lambda_{k,0}=\tau_{k,k}=\left\{
\begin{array}{ll}
1 \!-\! \sum\limits_{m=1}^M\lambda_{k,m}, &  k=0,\\
1\! -\! \sum\limits_{m=1}^{M}\lambda_{k,m} \!-\! \mu_k,  &  1 \leqslant k \leqslant K.
\end{array}
\right.
\end{align}
\end{theorem}
\vspace{-0.0cm}

In \figurename \ref{markov}, we present the Markov chain model with $M=2$. 
In each time slot, the queue length is increased no more than $M$ due to one new data arrival, while decreased by one due to one packet transmission. $\lambda_{k,0}$ is the probability that the queue length remains the same.

Matrix $\mathbf{\Lambda}$ is used to denote the $(K\!+\!1)$-by-$(K\!+\!1)$ transition probability matrix of the formulated Markov chain, in which, the $(j\!+\!1,i\!+\!1)$th element is $\tau_{i,j}$. Specifically, $\mathbf{\Lambda}$ is given by
\begin{displaymath}
\mathbf{\Lambda}=
\left[
\begin{array}{cccccccc}
\lambda_{0,0} & \mu_1 \\
\lambda_{0,1} & \lambda_{1,0} & \mu_2 \\
\lambda_{0,2} & \lambda_{1,1} & \lambda_{2,0} & \mu_3 \\
              & \lambda_{1,2} & \lambda_{2,1} & \lambda_{3,0} & \mu_4 \\
                                             &         &  \ddots       & \ddots   &  \ddots &\mu_K\\
                              &               &               & \lambda_{K\!-\!2,2} & \lambda_{k\!-\!1,1} & \lambda_{K,0}
\end{array}
\right].
\end{displaymath}

Let $\pi_{k}$ denote the steady-state probability of the queue state being in $k$. The stationary distribution of the Markov chain is denoted by $\bm{\pi} = [\pi_0,\pi_1,\cdots, \pi_K]^T$, where the superscript $T$ denotes the matrix transpose. Vectors $\mathbf{1}$ and $\mathbf{0}$ are used to denote the $(K+1)$-dimensional column vectors all of whose entries are zero and one, respectively. According to the property of the Markov chain, we have $\mathbf{\Lambda}\bm{\pi}=\bm{\pi}$ and $\mathbf{1}^T \bm{\pi} =1$. Hence, the stationary distribution $\bm{\pi}$ is the solution to the following linear equations
\begin{align}\label{steady-pi}
\Big[
\begin{array}{c}
\mathbf{Q}\\
\mathbf{1}^T
\end{array}
\Big]\bm{\pi} =
\Big[
\begin{array}{c}
0\\
1
\end{array}
\Big]
\end{align}
where the generator matrix $\mathbf{Q}$ is given by $(\mathbf{\Lambda}-\mathbf{I})$. From Eq. \eqref{steady-pi} and Theorem \ref{theorem1}, we can see that the steady-probability $\bm{\pi}$ is determined by the scheduling parameters, i.e., the transmission probabilities.

\vspace{0.0cm}
\section{Delay and Power Analysis}\label{sec4}
\vspace{0.0cm}

In this section, we analyze the two fundamental performance metrics of our considered system, i.e., the average queueing delay and average power consumption.

When the stationary distribution of the Markov chain is obtained, the average queue length is expressed as $E\{q[n]\}= \sum_{k=0}^{K} k\pi_k$.
Then, according to the Little's Law \cite{kleinrock1975queueing}, the average queueing delay is obtained as
\begin{align}\label{eq10}
D=\frac{1}{\bar{a}}\sum_{k=0}^{K} k\pi_k.
\end{align}

We use $c[n]$ to denote the transmission power in the $n$th time slot. Based on the proposed scheduling scheme in Section \ref{sec3}, we have $c[n]=P_w$ and $c[n]=0$, respectively, when one packet is transmitted over the channel state $w$ and no transmission takes place. For convenience, let us set $P_0=0$. Let $\psi_{k,w}$ denote the probability of $c[n]=P_w$ $(0\leqslant w \leqslant W)$. 
\begin{lemma}\label{lemma1} The conditional probability $\psi_{k,w}$ are expressed as
\begin{align}\label{eq11}
\psi_{k,w} =& \text{Pr} \{ c[n]\!=\!P_w \big| q[n\!-\!1]\!=\!k,h[n]\!=\! 'state\  w' \}  \nonumber \\
       = & \left\{
\begin{array}{ll}
\sum_{m=0}^{M} \theta_m f_{k\!+\!m,w}, & 1\leqslant w \leqslant W, \\
\! 1\!-\!\sum_{w=1}^{W} \psi_{k,w}, & w = 0.
\end{array}
\right.
\end{align}
\end{lemma}


With the conditional probability given by Eq. \eqref{eq11}, the average power consumption is obtained as

\begin{align}\label{eq12}
\bar{P}= & \sum_{k\!=\!0}^K \sum_{w\!=\!1}^W  \text{Pr}\{q[n\!-\!1]=k\}\text{Pr}\{h[n]='state\  w'\} \nonumber\\
& \times \text{Pr} \{ c[n]\!=\!P_w \big| q[n\!-\!1]\!=\!k,h[n]\!='state\  w' \} \cdot P_w \nonumber\\
=  &\sum_{k=0}^K \sum_{w=1}^W  \pi_k \eta_w \psi_{k,w}P_w  \\
= & \sum_{k=0}^K  \pi_k \sum_{w=1}^W  \eta_w P_w \sum_{m=0}^M   \theta_m f_{k\!+\!m,w}. \nonumber
\end{align}

\section{Optimal Delay-Power Tradeoff} \label{sec5}

To find the optimal scheduling probabilities $\{f_{k+m,w} | ,0 \leqslant k \leqslant K, 0 \leqslant m \leqslant M, 1\leqslant w \leqslant W \}$, optimization problems are formulated to minimize the average queueing delay $D$ given the power constraint $P_{aver}$.

\subsection{Optimization Problem}

Based on the analyses in Sections \ref{steady-proc} and \ref{sec4}, we know that the steady-probability $\bm{\pi}$ and the power consumption are determined by the scheduling probabilities $\{f_{k+m,w}\}$. To find the optimal scheduling probabilities, we formulate an optimization problem as follows:
\begin{equation}\label{tradeoff1}
\begin{split}
&\mathop{\text{min}}\limits_{\{f_{k+m,w} \}} \  D=\frac{1}{\bar{a}}\sum_{k=0}^{K} k\pi_k \\
&\text{s.t.}
 \left\{
\begin{array}{ll}
\bar{P}  \leqslant P_{aver}   & (a)\\
\mathbf{Q}\bm{\pi}=0 &(b)\\
\mathbf{1}^T \bm{\pi} =1  & (c)\\
 \bm{0} \preceq \bm{\pi} \preceq \bm{1} &(d)\\
f_{k+m,w} \in [0,1],  & (e)
\end{array}
\right.
\end{split}
\end{equation}
where $"\preceq"$ represents component-wise inequality between vectors and $0 \leqslant k \leqslant K$, $0 \leqslant m \leqslant M$, $1\leqslant w \leqslant W$. In \eqref{tradeoff1}, constraint (\ref{tradeoff1}.a) denotes the maximum power constraint. Constraints (\ref{tradeoff1}.b-\ref{tradeoff1}.d) are derived from the properties of the Markov chain. Constraint (\ref{tradeoff1}.d) shows the range of steady-state probability. Constraint (\ref{tradeoff1}.e) indicates the range of the variables $\{f_{k+m,w}\}$. Notice that, the problem \eqref{tradeoff1} is a non-linear problem optimization problem with $\{ f_{k+m,w} \}$ being its variables. It's rather difficult to solve it to obtain the optimal solution $\{ f_{k+m,r}^* \}$. To make it tractable, we convert optimization problem $\eqref{tradeoff1}$ into an equivalent LP problem via variable substitution.

\subsection{Formulation of Linear Programming Problem}

To obtain an LP optimization problem, we introduce a set of new variables $\{y_{k,w} | 0 \leqslant k \leqslant K, 1\leqslant w \leqslant W \}$ as
\begin{align}\label{eq13}
\vspace{-0.0cm}
y_{k,w}& = \sum\limits_{m=0}^{M} \pi_{k\!+\!1\!-\!m}  \theta_{m} f_{(k+1-m)+m,w} \nonumber \\
       & = \sum\limits_{m=0}^{M} \pi_{k\!+\!1\!-\!m}  \theta_{m} f_{k+1,w},\ \
\vspace{-0.0cm}
\end{align}
In Eq. \eqref{eq13}, $\pi_{k+1-m} \theta_{m} f_{k+1,w}$ is the probability of transmitting one packet, i.e., $s[n]=1$, when $q[n]=k\!+\!1-\!m$ and $a[n]=m$. Thus, $y_{k,w}$ is the probability that there are $k$ packets backlogged in the queue after one packet is transmitted over channel state $w$.
\begin{lemma}\label{lemma2}
From Eq. \eqref{eq13}, the average queue length and the average power consumption can be transformed as
\begin{equation}
\begin{split}
\left\{
\begin{array}{ll}
D=\frac{1}{\bar{a}^2} \Big( \sum\limits_{k=0}^{K} \sum\limits_{w=1}^W k \eta_w y_{k,w} - \xi \Big) \\
P=\sum\limits_{k=0}^{K}\sum\limits_{w=1}^W \eta_w P_w y_{k,w},
\end{array}
\right.
\end{split}
\end{equation}
where $\xi = \sum_{m=1}^{M-1} \frac{m(m+1)}{2}\theta_{m+1}$ is a constant.
\end{lemma}

\begin{lemma}\label{lemma3}
Constraints (\ref{tradeoff1}.b,c,e) 
can be converted into the following constraints respectively.
\begin{align}
\left\{
\begin{array}{ll}\label{new constraint}
\sum\limits_{i=0}^{M-1}\pi_{k-i} r_i = \sum\limits_{w=1}^W \eta_w y_{k,w},  &(a)  \\
\sum\limits_{k=0}^{K}\sum\limits_{w=1}^W \eta_w y_{k,w} = \bar{a},  &(b)   \\
0 \leqslant y_{k,w} \leqslant \sum\limits_{m=0}^{M} \theta_{m} \pi_{k\!+\!1\!-\!m}, &(c)
\end{array}
\right.
\end{align}
where  $r_i= \sum\limits_{m=i\!+\!1}^M \!\! \theta_m$.
\end{lemma}

From (\ref{new constraint}.a), we introduce a $(K\!+\!1) \!\times\! \big[W(K\!+\!1)\big]$-dimensional matrix $\bm{G}$, in which, the $(k+1)$th row is denoted by $\bm{g}_{k+1}$ and given as
\begin{align}
\left\{
\begin{array}{cll}
\bm{g}_{k+1} &= \frac{1}{r_0} \bm{l}_{1}, & k=0,   \\
\bm{g}_{k+1} &= \frac{1}{r_0} ( \bm{l}_{k+1} - \!\! \sum\limits_{i=1}^{M-1} r_i \bm{g}_{k-i}), & 1 \leqslant k \leqslant K. \label{GGneration}
\end{array}
\right.
\end{align}
In Eq. \eqref{GGneration}, $\bm{l}_{k+1}$ is a $W(K\!+\!1)$-dimensional vector, in which, the $(Wk+w)$th element is $\eta_w$ while other elements are zero. In this way, the steady-state probability can be linearly expressed by the set of parameters ${y_{k,w}}$
\begin{align}\label{pi-y}
\pi_k = \sum_{i=0}^{K}\sum_{j=1}^{W}  G_{(k+1,iW+j)} \cdot y_{i,j},
\end{align}
where $G_{(i,j)}$ is the $(i \times j)$th element of matrix $\bm{G}$.


\begin{theorem}\label{the3}
The optimization problem \eqref{tradeoff1} is equivalent to the following LP problem.
\begin{equation}\label{tradeoff2}
\begin{split}
&\mathop{\text{min}}\limits_{\{ y_{k,w} \}} \  D=\frac{1}{\bar{a}^2} \Big( \sum\limits_{k=0}^{K}  \sum\limits_{w=1}^W k \eta_w y_{k,w} - \xi \Big) \\
& \!\!\!\! \text{s.t.}\!
 \left\{\!\!
\begin{array}{ll}
\bar{P}=\sum\limits_{k=0}^{K}\sum\limits_{w=1}^W \eta_w P_w y_{k,w}  \leqslant P_{aver}  \!\! & (a)\\
\sum\limits_{k=0}^{K}\sum\limits_{w=1}^W \eta_w y_{k,w} = \bar{a}    \!\!  &   (b)\\
0 \! \leqslant \! y_{k,w} \!\! \leqslant \! \sum\limits_{m=0}^{M}  \! \theta_{m} \! \sum\limits_{i=0}^{K} \! \sum\limits_{j=1}^{W} \! G_{(k+2-m,iW+j)}\! \cdot \! y_{i,j}    \!\! & (c)\\
\end{array}
\right.
\end{split}
\end{equation}
\end{theorem}
\vspace{-0.3cm}

\vspace{0.0cm}
\subsection{Optimal Scheduling Policy}

\begin{lemma}
The optimal solution $y_{k,w}^*$ to LP problem \eqref{tradeoff2} has a threshold structure described as
\begin{equation}\label{solution}
\begin{split}
y_{k,w}^*=\left\{
\begin{array}{ll}
0,   & k < K_w^*-1,\\
\sum_{m=0}^{M} \theta_{m} \pi_{k+1-m}^*,  & k > K_w^*-1,
\end{array}
\right.
\end{split}
\end{equation}
where $K_w^*$ is the threshold imposed on the queue length in channel state $w$. 
\end{lemma}
Comparing the optimal solution in Eq. \eqref{solution} with the definition of $y_{k,w}$ given in Eq. \eqref{eq13}, we obtain the optimal scheduling parameters described in the following theorem.
\begin{theorem}
The optimal scheduling policy corresponds to a threshold-based policy. The optimal scheduling probabilities are given by
\begin{equation}\label{eq14}
\begin{split}
f_{k,w}^* = \left\{
\begin{array}{ll}
0, k < K_w^*,\\
1, k > K_w^*,
\end{array}
\right.
\end{split}
\end{equation}
and $f_{K_w^*,w}^* = y_{K_w^*-1,w}^* (\sum_{m=0}^{M} \pi_{k\!+\!1\!-\!m}^*  \theta_{m})^{-1}$. 
\end{theorem}

From Eq. \eqref{eq13}, we know $f_{k+1,w}^*= f_{(k+1-m)+m,w}^*$, that is, the optimal scheduling probability is based on the total number of backlogged packets, which is updated after each packet arrival. Specifically, at the beginning of each slot, the scheduler collects the information of the channel state $h[n]='state\  w'$, queue state $q[n-1]$ and data arrival state $a[n]$, then makes a decision of transmitting one packet if $q[n-1]+a[n]>K_w^*$, or keeping silence if $q[n-1]+a[n]<K_w^*$, or delivering one packet with probability $f_{K_w^*,w}^*$ when $q[n-1]+a[n]=K_w^*$.

It is not a trivial work to obtain a closed-form expression for the thresholds $K_w^*$. Fortunately, we can adopt many low-complexity mature algorithms to solve the LP problem \eqref{tradeoff2}. Based on the optimal solution to \eqref{tradeoff2}, we can obtain the thresholds $K_w^*$ numerically. Hence, we still obtain the optimal threshold-based policy which simply depends on the optimal threshold on the queue state for each channel state. Based on the optimal solution, we derive the threshold-based policy as the optimal scheduling policy for delay-power tradeoff.

\vspace{0.0cm}
\section{Numerical Results} \label{sec6}
\vspace{0.0cm}
%
%

\begin{figure*}[htb]
\centering
\captionsetup[subfloat]{labelformat=simple,captionskip=6bp,nearskip=6bp,farskip=0bp,topadjust=0bp}
\renewcommand{\thesubfigure}{(\alph{subfigure})}
\subfloat[Different arrival rates]{
\includegraphics[width=0.65\columnwidth]{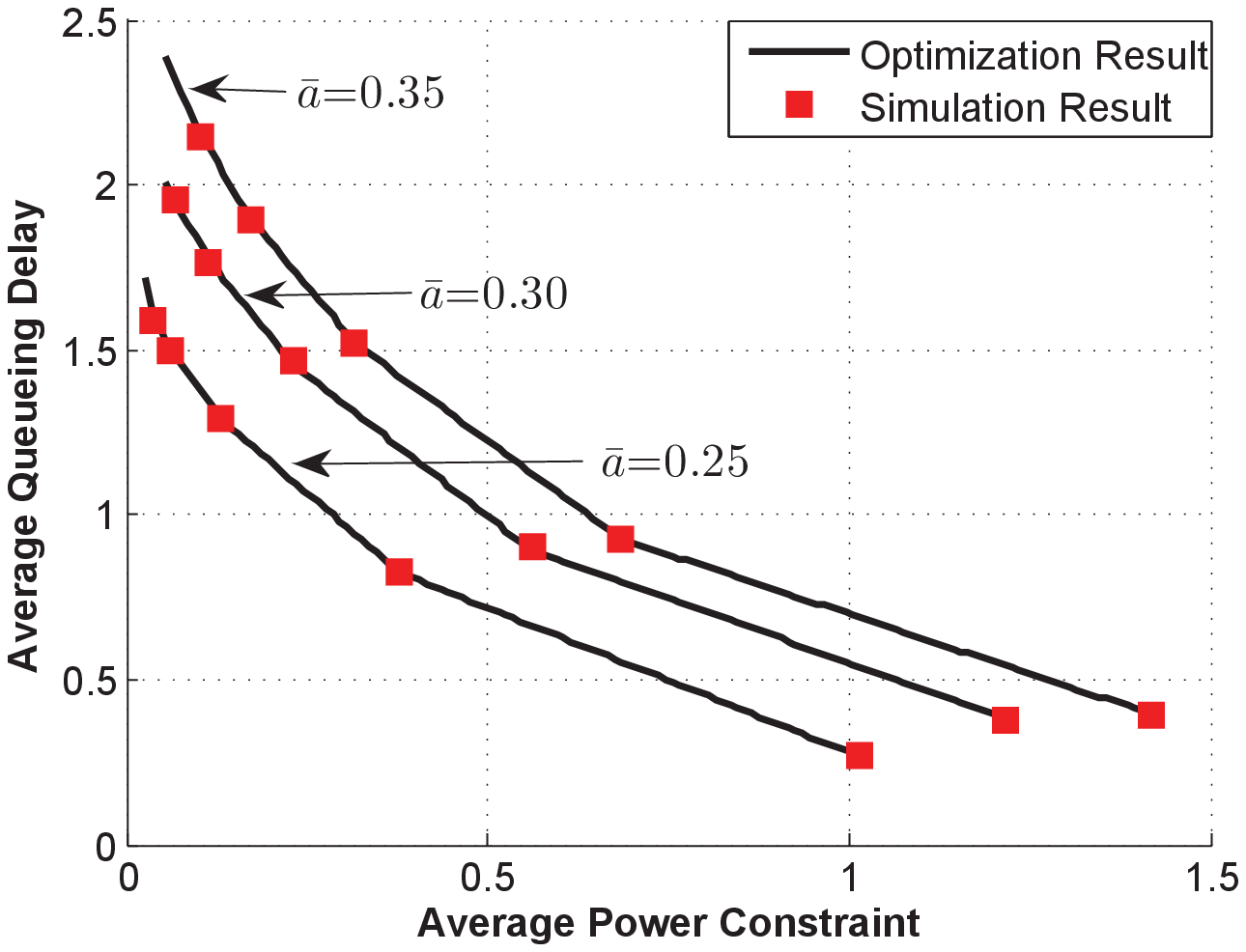} \label{sim}}
\hspace*{-0.3cm}
\subfloat[Different arrival variances]{
\includegraphics[width=0.7\columnwidth]{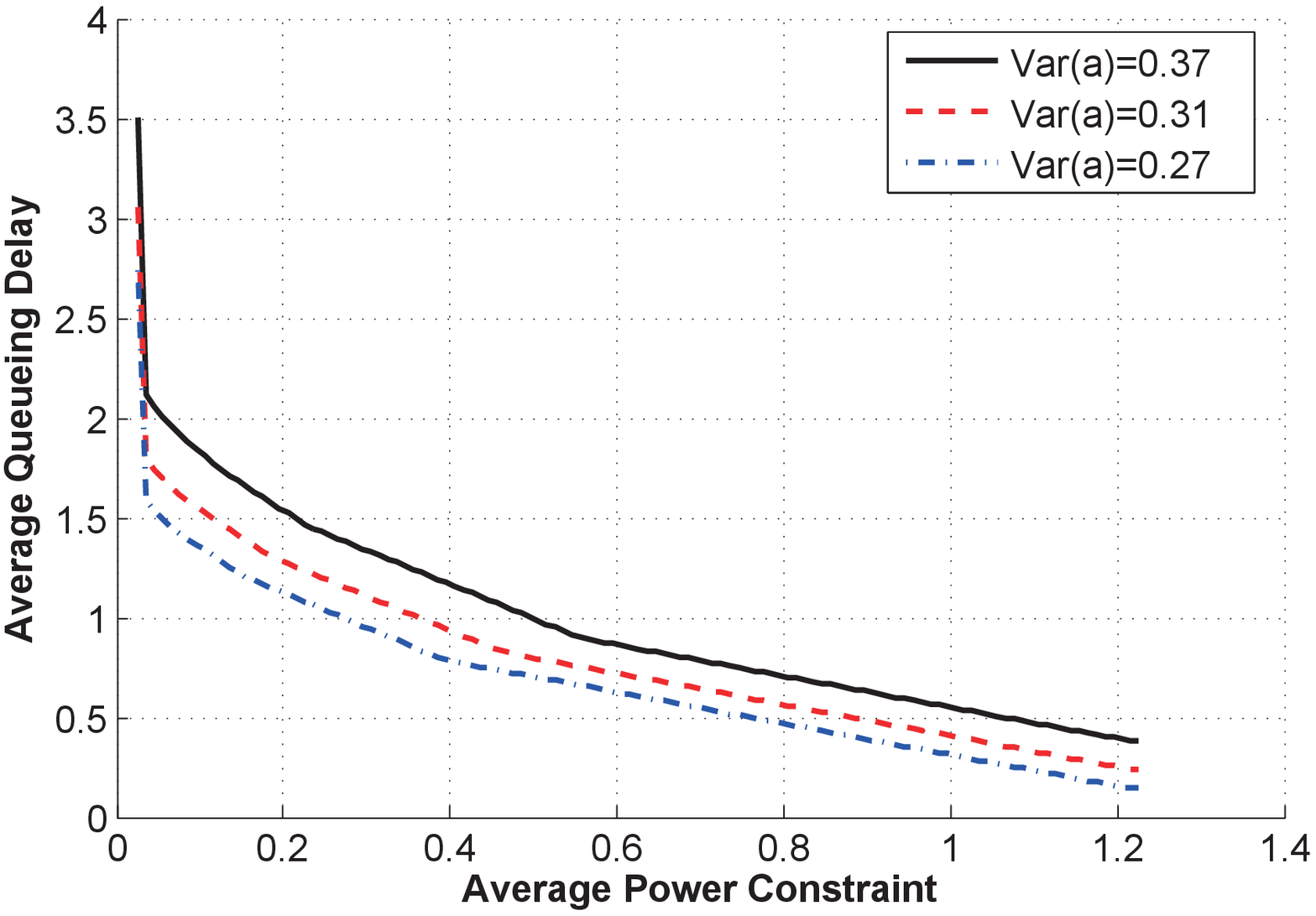} \label{sim_2}}
\hspace*{-0.3cm}
\subfloat[Different BER requirements]{
\includegraphics[width=0.67\columnwidth]{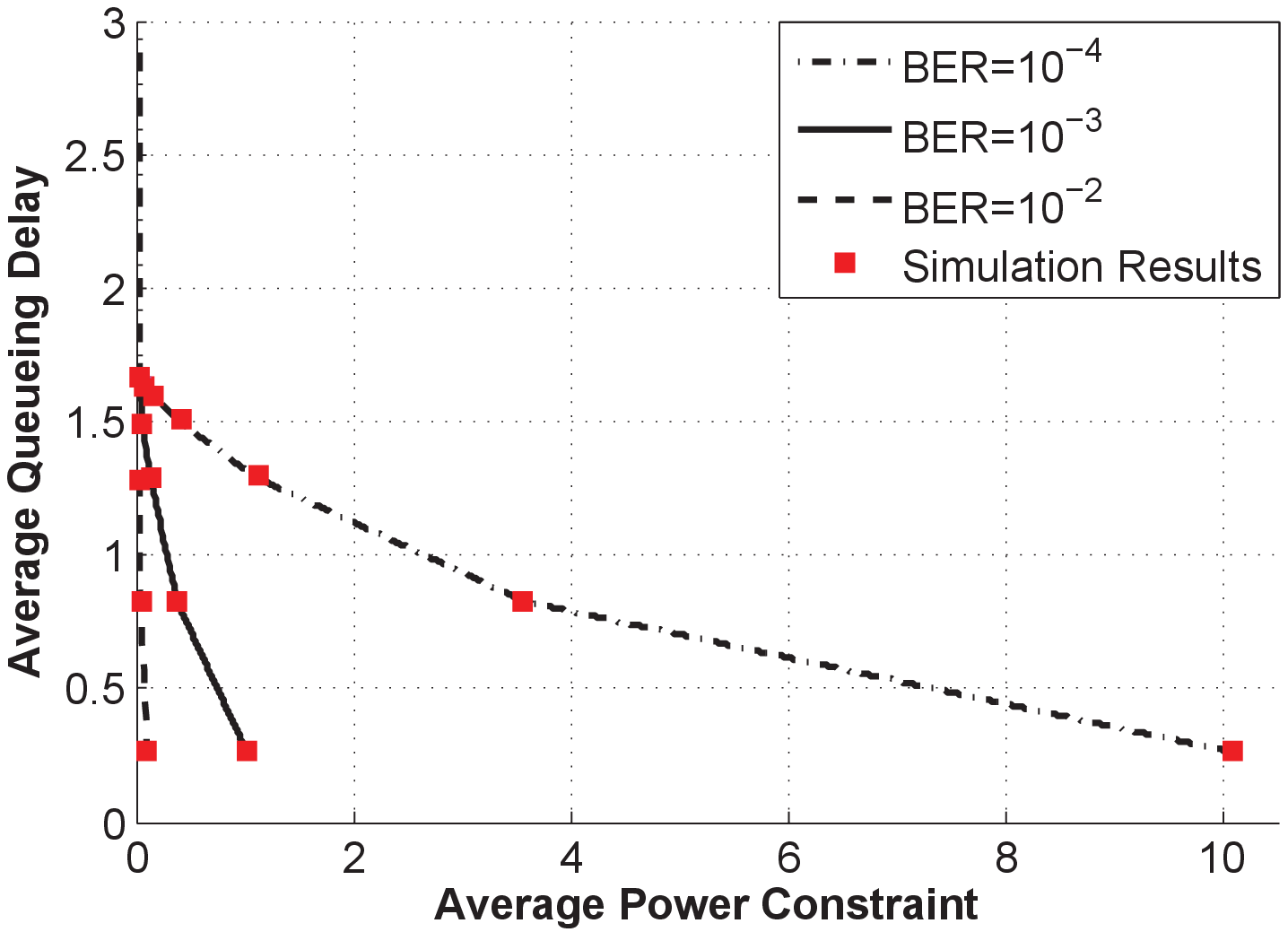} \label{sim_1} }
\caption{Optimal delay-power tradeoff curves for different scenarios}
\label{model}
\vspace{-0.0cm}
\end{figure*}


In this section, simulation results are given to validate the threshold-based policy demonstrate its potential. In the simulation, data packets are generated following a given probabilistic distribution $\{ \theta_{(\cdot)} \}$ and the maximum size of the arrival packet is limited by $M=2$. We adopt $W$-state block fading channel model with $W=4$. The probabilistic distribution of the $\eta_w$ which is defined as the probability of \{ $h[n]='state\  w'$\} is given by $[\eta_1, \eta_2, \eta_3, \eta_4] = [0.135, 0.232, 0.239, 0.394]$. The specific distributions of packet arrival and the transmission power can be found in Table \ref{simulation} for different simulations for convenience. Packets are delivered based on threshold-based policy. Each simulation runs over $10^7$ time slots. The theoretical results are plotted by lines (solid or dashed) while the simulation results are marked by red square dots.
\begin{table}[h]
\centering
\begin{tabular}{|p{0.3cm}|p{0.2cm}p{0.3cm}|p{0.45cm}|p{0.2cm}p{0.2cm}p{0.2cm}p{0.57cm}|p{0.18cm}|p{0.35cm}|p{0.3cm}p{0.3cm}|p{0.3cm}p{0.3cm}p{0.3cm}|}  
\hline
\multicolumn{3}{|c|}{（Arrival Rates）}&\multicolumn{5}{|c|}{（ BER Requirements ）}  & \multicolumn{4}{|c|}{（ Arrival Variances ）}\\
\hline
$\bar{a}$ & $\theta_1$ & $\theta_2$  & BER & $P_1$ & $P_2$ & $P_3$ & $P_4$ & $\bar{a}$ & Var  & $\theta_1$ & $\theta_2$\\ \hline  
0.25 &0.15 &0.05  &$10^{-4}$ &0.08 &0.11 &0.24 &101.64 & 0.3 & 0.27  &0.24 &0.03  \\ \hline        
0.30  &0.14 &0.08  &$10^{-3}$ &0.04 &0.08 &0.16 &10.14 & 0.3 & 0.31 &0.20 &0.05 \\ \hline       
0.35 &0.17 &0.09  &$10^{-2}$ &0.02 &0.04 &0.08 &0.99  & 0.3  & 0.37 &0.14 &0.08  \\ \hline
\end{tabular}
\caption{Simulation parameter settings}\label{simulation}
\end{table}

In \figurename\ref{sim}, the transmission power $P_w$ is adopted to meet BER requirement $10^{-3}$. We plot the delay-power tradeoff curves under different packet arrival rates, i.e., $\bar{a}$ is set to $0.25$, $0.3$, and $0.35$, respectively. We can see that the theoretical results are in good agreement with the simulation results. The delay-power tradeoff curve is piecewise linear in accordance to the threshold-based policy. Besides, the average delay decreases when the average power increases, and when the available power decreases and approaches zero, the queueing delay will increase dramatically and grow to infinity since the queueing system is unstable. Given the same power constraint, the queueing delay is larger for a higher packet arrival rate since more packet will be needed to be transmitted.

In \figurename\ref{sim_2}, the BER requirement is set to $10^{-3}$. We present the delay and power tradeoffs under different variances of $\theta_{m}$ when the average arrival rate is fixed to be $\bar{a}=0.3$. From \figurename\ref{sim_2}, it is observed that higher queueing delay is induced when the data arrival variance is larger in the case with the identical average data arrival rate. Due to higher busty arrivals, some packets have to wait for longer time before they are transmitted, which as a result, leads to a larger queueing delay.

In \figurename\ref{sim_1}, the delay-power tradeoff curves are plotted in the scenarios that $\bar{a}$ is set to be $0.3$, and the target BER is set to $10^{-4}$, $10^{-3}$ and $10^{-2}$, respectively. For a higher BER requirement, the scheduler has to use more power to transmit one packet for the same channel quality to make sure the packet can be received successfully. Thus, higher queueing delay will be if the power consumption is identical but the BER requirement is more strict.

%


\vspace{0.0cm}
\section{Conclusion} \label{sec7}
\vspace{0.0cm}

In this paper, we studied the power-constrained delay-optimal scheduling problem in wireless systems, where an arbitrary packet arrival pattern and multi-state block-fading channels were considered. A probabilistic scheduling policy was proposed to schedule data transmissions over $W$-state wireless fading channel based on the queue length and the channel state. 
The average queueing delay and average power consumption were analysed based on Markov reward process. Based on this, optimization problem was formulated to capture the delay-power tradeoff by optimizing the scheduling probabilities. Theoretical analysis revealed the structure of the optimal solution. We obtained the optimal threshold-based policy accordingly. It is found that an optimal transmission policy naturally takes action in line with the queue length and the channel state. It always seeks to exploit a good channel in the mean time keep the queue length short. At a result, thresholds are imposed on the queue for different channel states. Specifically, if the queue length is higher than the threshold, the scheduler should transmit to decrease the latency. Otherwise, it should keep silent to improve the energy efficiency.
\vspace{0.0cm}

\bibliographystyle{IEEEtran}
\bibliography{IEEEabrv,scheduling}

\begin{thebibliography}{10}
\providecommand{\url}[1]{#1}
\csname url@samestyle\endcsname
\providecommand{\newblock}{\relax}
\providecommand{\bibinfo}[2]{#2}
\providecommand{\BIBentrySTDinterwordspacing}{\spaceskip=0pt\relax}
\providecommand{\BIBentryALTinterwordstretchfactor}{4}
\providecommand{\BIBentryALTinterwordspacing}{\spaceskip=\fontdimen2\font plus
\BIBentryALTinterwordstretchfactor\fontdimen3\font minus
  \fontdimen4\font\relax}
\providecommand{\BIBforeignlanguage}[2]{{%
\expandafter\ifx\csname l@#1\endcsname\relax
\typeout{** WARNING: IEEEtran.bst: No hyphenation pattern has been}%
\typeout{** loaded for the language `#1'. Using the pattern for}%
\typeout{** the default language instead.}%
\else
\language=\csname l@#1\endcsname
\fi
#2}}
\providecommand{\BIBdecl}{\relax}
\BIBdecl

\bibitem{6824752}
J.~Andrews, S.~Buzzi, W.~Choi, S.~Hanly, A.~Lozano, A.~Soong, and J.~Zhang,
  ``What will 5g be?'' \emph{IEEE Journal on Selected Areas in Communications},
  vol.~32, no.~6, pp. 1065--1082, June 2014.

\bibitem{collins1999transmission}
B.~Collins and R.~L. Cruz, ``Transmission policies for time varying channels
  with average delay constraints,'' in \emph{Proc. Allerton Conf.
  Communication, Control, and Computing, Monticello, IL}, 1999, pp. 709--717.

\bibitem{995554}
R.~A. Berry and R.~G. Gallager, ``Communication over fading channels with delay
  constraints,'' \emph{IEEE Transactions on Information Theory}, vol.~48,
  no.~5, pp. 1135--1149, May 2002.

\bibitem{berry2004cross}
R.~A. Berry and E.~M. Yeh, ``Cross-layer wireless resource allocation,''
  \emph{IEEE Signal Process. Mag.}, vol.~21, no.~5, pp. 59--68, 2004.

\bibitem{ata2005dynamic}
B.~Ata, ``Dynamic power control in a wireless static channel subject to a
  quality-of-service constraint,'' \emph{Oper. Res.}, vol.~53, no.~5, pp.
  842--851, 2005.

\bibitem{uysal2002energy}
E.~Uysal-Biyikoglu, B.~Prabhakar, and A.~El~Gamal, ``Energy-efficient packet
  transmission over a wireless link,'' \emph{IEEE/ACM Trans. Netw.}, vol.~10,
  no.~4, pp. 487--499, 2002.

\bibitem{zafer2009calculus}
M.~A. Zafer and E.~Modiano, ``A calculus approach to energy-efficient data
  transmission with quality-of-service constraints,'' \emph{IEEE/ACM Trans.
  Netw.}, vol.~17, no.~3, pp. 898--911, 2009.

\bibitem{4533700}
J.~Yang and S.~Ulukus, ``Delay-minimal transmission for energy constrained
  wireless communications,'' in \emph{2008 IEEE International Conference on
  Communications}, May 2008, pp. 3531--3535.

\bibitem{altman1999constrained}
E.~Altman, \emph{Constrained Markov decision processes}.\hskip 1em plus 0.5em
  minus 0.4em\relax CRC Press, 1999, vol.~7.

\bibitem{hernandez2012discrete}
O.~Hern{\'a}ndez-Lerma and J.~B. Lasserre, \emph{Discrete-time Markov control
  processes: basic optimality criteria}.\hskip 1em plus 0.5em minus 0.4em\relax
  Springer Science \& Business Media, 2012, vol.~30.

\bibitem{4567575}
M.~Goyal, A.~Kumar, and V.~Sharma, ``Optimal cross-layer scheduling of
  transmissions over a fading multiaccess channel,'' \emph{IEEE Transactions on
  Information Theory}, vol.~54, no.~8, pp. 3518--3537, Aug 2008.

\bibitem{chen2007optimal}
W.~Chen, Z.~Cao, and K.~B. Letaief, ``Optimal delay-power tradeoff in wireless
  transmission with fixed modulation,'' in \emph{Proc. IEEE IWCLD}, 2007, pp.
  60--64.

\bibitem{7417380}
M.~Wang and W.~Chen, ``Achieving the optimal delay-power tradeoff in wireless
  transmission with arbitrarily random packet arrival: A cross-layer
  approach,'' in \emph{2015 IEEE Global Communications Conference (GLOBECOM)},
  Dec 2014, pp. 1--6.

\bibitem{qiao2009impact}
D.~Qiao, M.~C. Gursoy, and S.~Velipasalar, ``The impact of qos constraints on
  the energy efficiency of fixed-rate wireless transmissions,'' \emph{IEEE
  Transactions on Wireless Communications}, vol.~8, no.~12, pp. 5957--5969,
  2009.

\bibitem{kleinrock1975queueing}
L.~Kleinrock, ``Queueing systems, volume 1: Theory,'' \emph{Lecture Notes in
  Computer Science}, 1975.

\end{thebibliography}
\balance

%

\end{document}